\documentclass[twocolumn,letterpaper]{IEEEAerospaceCLS}  % only supports two-column, letterpaper format
\usepackage{amsmath,amssymb}
\usepackage{lscape}
\usepackage{comment}
\usepackage{leftidx}
\usepackage{graphicx}
\usepackage{booktabs}
\usepackage{afterpage}
\usepackage{algorithm}
\usepackage{float}
\usepackage{caption}
\usepackage{subcaption}
\usepackage{wrapfig}
\usepackage{enumerate}
\usepackage{calc}
\usepackage{fp}
\usepackage{tikz}
\usepackage{epstopdf}
\usepackage{siunitx}
\usepackage{enumerate}
\usepackage{layouts}
\usepackage{algpseudocode}
\usepackage{pifont}
\usepackage{multirow}
\usepackage{adjustbox}
\usepackage{import}
\usepackage{caption}
\usepackage{subcaption}

\usepackage[]{graphicx}    % We use this package in this document
\newcommand{\ignore}[1]{}  % {} empty inside = %% comment

\begin{document}
\title{Multi-Stage Fusion Architecture for Small-Drone Localization and Identification Using Passive RF and EO Imagery: A Case Study}

\author{%
Thakshila Wimalajeewa Wewelwala*, Thomas W. Tedesso  and Tony Davis\\
Air and Missile Defense Sector (AMDS)\\
Johns Hopkins University Applied Physics Laboratory (JHU/APL) \\
 11100 Johns Hopkins Rd, Laurel, MD 20723\\
*Thakshila.Wimalajeewa@ieee.org
%\and
%Tom Tedesso\\
%Department of ECE\\
%University of Nowhere\\
%Nowhere, ZS 99999\\
%jane.smith@nowhere.edu
%%%% IMPORTANT: Use the correct copyright information--IEEE, Crown, or U.S. government. %%%%%
%\thanks{\footnotesize}              % This creates the copyright info that is the correct 2024 data.
\thanks{{DISTRIBUTION STATEMENT A: Approved for public release: distribution is unlimited}}
\thanks{\footnotesize 979-8-3503-0462-6/24/$\$31.00$ \copyright2024 IEEE}
        % Use this copyright notice only if you are employed by the U.S. Government.
%\thanks{{979-8-3503-0462-6/24/$\$31.00$ \copyright2024 Crown}}          % Use this copyright notice only if you are employed by a crown government (e.g., Canada, UK, Australia).
%\thanks{{979-8-3503-0462-6/24/$\$31.00$ \copyright2024 European Union}}    % Use this copyright notice is you are employed by the European Union.
}

\maketitle

\thispagestyle{plain}
\pagestyle{plain}

\maketitle

\thispagestyle{plain}
\pagestyle{plain}

\begin{abstract}
Reliable detection, localization and identification of small drones is essential to promote safe, secure and privacy-respecting operation of Unmanned-Aerial Systems (UAS), or simply, drones. This is an increasingly  challenging problem with only single modality sensing, especially, to detect and identify small drones. In this work, a multi-stage fusion architecture using passive radio frequency (RF)  and electro-optic (EO) imagery data is developed to leverage the synergies of  the modalities to improve the overall tracking and classification  capabilities.  For detection with  EO-imagery, supervised deep learning based techniques as well as unsupervised foreground/background separation techniques are explored to cope with challenging environments. Using real collected data for Group 1 and 2  drones, the capability of each algorithm is quantified. In order to compensate for any performance gaps in detection with only EO imagery  as well as to provide a unique device identifier for  the  drones, passive RF is integrated with EO imagery whenever available.  In particular, drone detections in the image plane are combined with passive RF location estimates via detection-to-detection association after 3D to 2D transformation. Final tracking is performed on the composite detections in the 2D image plane. Each track centroid is given a unique identification  obtained via RF fingerprinting. The proposed fusion architecture is tested and the tracking and performance is quantified over the range to  illustrate the effectiveness of the proposed approaches using  simultaneously collected passive RF and EO data at the Air Force Research Laboratory (AFRL) through ESCAPE-21 (Experiments, Scenarios, Concept of Operations, and Prototype  Engineering)  data collect.
\end{abstract}

\tableofcontents

%%%%%%%%%%%%%%%%%%%%%%%%%%%%%%%%%%%%%%
\section{Introduction}
%%%%%%%%%%%%%%%%%%%%%%%%%%%%%%%%%%%%%%
While unmanned aerial systems (UASs) or more specifically, drones,   are being actively used for a variety of civil and commercial applications including photography,   package delivery, rescue operations, wireless communication,  and sheer enjoyment of flight, the threat of small UASs (sUASs) to national security is increasing  \cite{Humphreys_2015,Solodov_18,Fotouhi_19}. Based on the payload capability, the drones can be used for various illegal activities and even be used as dangerous weapons after loading them with explosive material \cite{Sayler_2018,Cubber_2019}.  In order to implement an effective counter-UAS (cUAS) system to prevent any security breaches  or undesirable impact on the public safety and national security, it is vital to detect and identify these small, in particular Group 1 and 2 \cite{Drone_class},  drones reliably and understand their intent. There are a variety of sensing modalities that can be used for drone surveillance including radar, electro-optical (EO), infrared (IR) imagery, passive radio frequency (RF), acoustic, etc..  Recent review papers present the advantages and disadvantages of the individual modalities for drone detection and identification \cite{Taha_19,Wang_21}.

\begin{figure*}
    \centering
   {\includegraphics[width=0.5\textwidth]{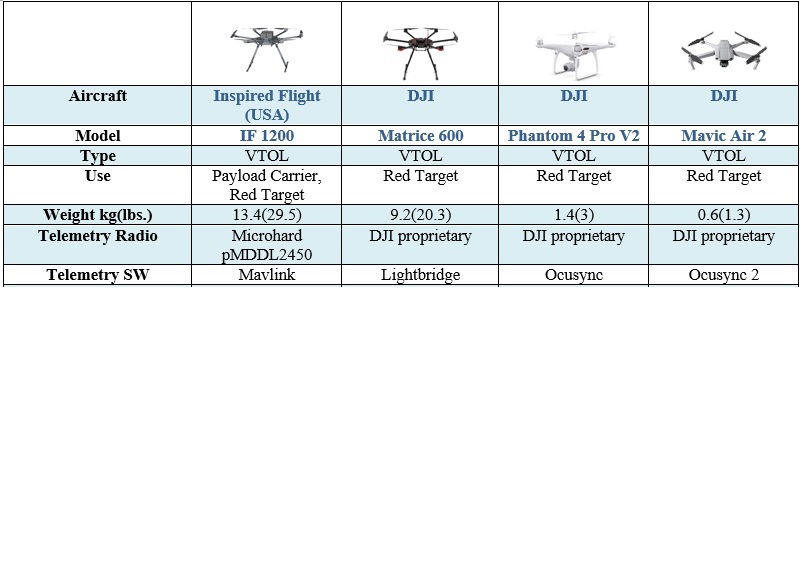}}
    \caption{AFRL ESCAPE2021 data collect:  Types of drones used as targets; picture credit: AFRL/RI [13]}
    \label{fig:Drone_spec}
\end{figure*}

As shown in multiple studies, the ability of reliable detection, classification and localization of  drones with only a single modality is quite challenging. The conventional radar (S and X band) suffers from  small radar cross section (RCS) of most of these small drones, thus,  the detection and discrimination capability is limited. To obtain higher range resolution, radar at higher frequencies (Ku/K/Ka/millimeter wave bands) could be used;  however, the maximum range that the drones can be detected/classified  is very limited.  For example, the authors in \cite{Kim_17} quantified the classification accuracy with micro-doppler imagery of a Ku band radar where the maximum distance considered is $\sim 100m$.  Video or imagery is another commonly used modality to detect and track drones; however, the performance depends on the drone size, availability of line of sight (LOS), and resemblance of other objects like birds. Drone detection and classification using  Deep Learning with EO imagery is an active research area, however, for small drones (Group 1, 2), reliable  detection is still challenging  even with the recent advances in deep learning when the drones occupy only a few pixels in the image pixel plane. On the other hand foreground/background separation algorithms to detect drones can perform poor in the presence of high clutter and slowly moving clouds, and also when the drones hover for a long time.  Most of the commercial drones designed for surveillance  and hobbyists use downlink transmission  to relay messages to a remote controller. These downlink signals can be used for detection, classification and localization of drones when available. Further,   passive RF (passive RF and P-RF, for passive RF, are used   interchangeably in the rest of the paper) can be used to infer useful information regarding the drone intent by analyzing the transmit waveform patterns and the RF fingerprint \cite{Ezuma_19,Barnard_2007,Kosolyudhthasarn_2018}.   Acoustic is another modality that can convey information regarding the drone payload via acoustic fingerprinting \cite{Kartashov_2021}. However, research on understanding  the full capabilities of acoustic modality for drone detection and identification  is still in its infancy.
\begin{figure}
    \centering
   {\includegraphics[width=0.4\textwidth]{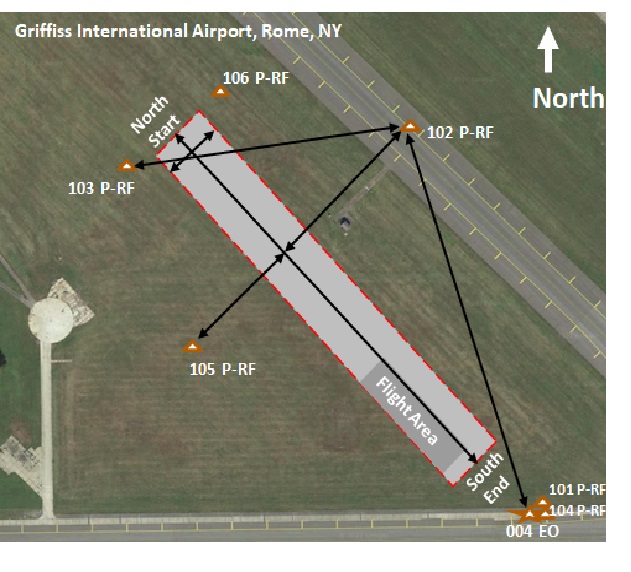}}
    \caption{AFRL ESCAPE2021 data collect: Sensor Layout with passive RF and EO Sensors; picture credit: AFRL/RI [13]}
    \label{fig:layout}
\end{figure}

The  goal of this paper is to evaluate the drawbacks of individual passive RF and EO modalities using collected data for Group 1, 2 drones and exploit the diversity across multiple modalities to design a reliable  fusion architecture for drone detection, identification  and tracking.  In this work, we consider how passive RF can be integrated with EO to boost the overall detection and identification performance of small drones. To detect the drones in the 2D image plane with EO,  supervised deep learning algorithms as well as  advanced foreground/background separation algorithms are investigated.   We show that the state-of-the-art deep learning techniques fail to reliably detect and identify  drones when the drones are far away (several hundreds of meters)  from the camera and occupy only few pixels in the  image plane.  When the drones appear to be point targets in the image plane, advanced foreground/background separation techniques can be more reliable in  detecting drones (as moving outliers in the image) than  deep learning techniques. In order to label the detected drones at the 'point target' stage in 2D image-pixel  plane, we incorporate device ID obtained via RF fingerprinting  using passive RF when available. RF fingerprinting is quite challenging for drones when they transmit similar waveforms; however, the anomalous behavior of the RF front-end  of the drone transmitter  provides a unique RF fingerprint for each drone which can be  learned using advanced deep learning techniques.  To map the EO detections with passive RF, we use time-difference-of-arrival (TDOA) based localization with passive RF in the sensor plane and project the 3D location estimates to the 2D image plane for detection-to-detection association. Finally, a Kalman filter is used on the composite EO and RF detections in the 2D image plane to form tracks for each detected drone along with the device labels.

\subsection{Description of Data and Drones}
In order to benchmark the detection and classification capabilities of the proposed architecture, we use simultaneously collected real passive RF and EO data for four small drones. This data is collected at the 2021 ESCAPE II Data Collection (termed as ESCAPE2021 in the rest of the paper) hosted by AFRL at the Griffiss International Airport, Rome, NY (GIA-RME) and the AFRL Stockbridge, NY Test Site (STS)   \cite{Zulch_23}. Multi-modal data is collected for multiple target types  (ground vehicles, drones and dismounts) with a variety of sensor types including radar, passive RF, EO, IR, seismic, and acoustic sensors over different days. In this work, we use data for drone targets collected at EO and passive RF sensors where  a target description is shown  in Fig. \ref{fig:Drone_spec} and the sensor layout is illustrated in Fig. \ref{fig:layout}. The drone targets include 3 DJI drones and one Inspired Flight (IF) 1200 as shown in Fig.  \ref{fig:Drone_spec}. Phantom and Mavic can be considered to belong to  Group 1, while m600 and IF1200 can fall into the Group 2 category.

More information about this dataset is provided in Section \ref{sec_performance}. Using this data, the goal  is to quantify  (i) the range from the camera that the  small drones can be reliably detected using the deep learning algorithms as well as advanced foreground/background separation algorithms using EO imagery, (ii) device ID accuracy of the four drones with AI-enabled RF fingerprinting, and (iii) the tracking performance of the  RF and EO  fusion architecture over the range from the EO camera.

%\subsection{Paper Organization}
%The paper is organized as follows. Section \ref{sec_overview} presents the problem formulation, EO and passive RF single modality operation and the proposed fusion architecture. In Section \ref{sec_performance}, the performance of drone detection with EO only, drone device ID performance with passive RF and combined track performance is evaluated using ESCAPE 2021 data. Section \ref{sec_conclusion} provides concluding remarks of the paper.
%%%%%%%%%%%%%%%%%%%%%%%%%%%%%%%%%%%%%%%%%%%
%\section{Background and Motivation}
%There are abundant of papers by now that investigates the use of deep learning  techniques for drone detection with EO imagery \cite{}. In --, the capabilities of fastRCNN
%
%\subsection{RF and EO Fusion}
%The proble of RF (mostly radar) and EO fusion is of interest to autonomous driving \cite{Aziz_Radar20}
%%%%%%%%%%%%%%%%%%%%%%%%%%%%%%%%%%%%%%%%%%%
\begin{figure}
    \centering
   {\includegraphics[width=0.4\textwidth]{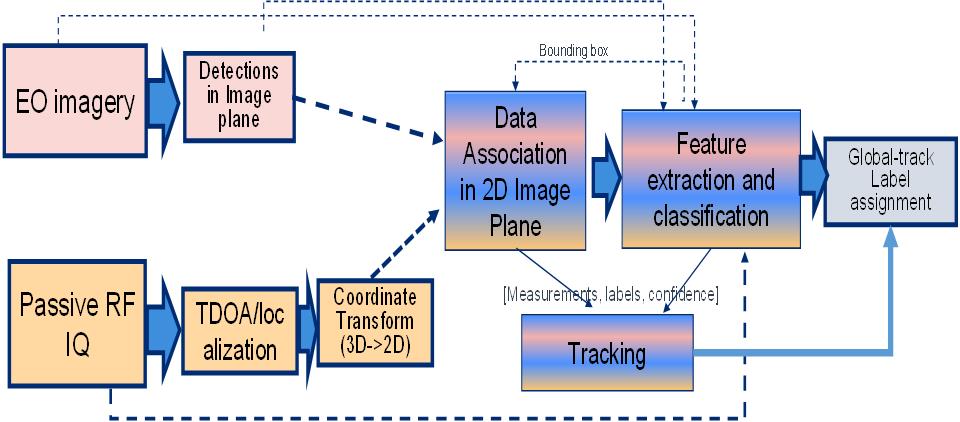}}
    \caption{EO and Passive RF individual operations and the fusion architecture for detection, identification and tracking of drones}
    \label{fig:Fusion_arch}
\end{figure}
\section{Problem Formulation/Solution Overview}\label{sec_overview}
%The goal of this work is to understand and quantify the individual capabilities of EO imagery and passive RF as well as fusion performance of the two modalities  for small drone detection, identification and tracking using real collected data.
The  proposed  fusion  architecture is  depicted in Fig. \ref{fig:Fusion_arch} along with individual operations of each modality.

\subsection{Drone Detection with EO Imagery}
To detect drones in EO imagery,  we explore both state-of-the-art deep learning based techniques as well as foreground/background separation techniques and illustrate the advantages of one over the other.
\begin{figure}
    \centering
   {\includegraphics[width=0.4\textwidth]{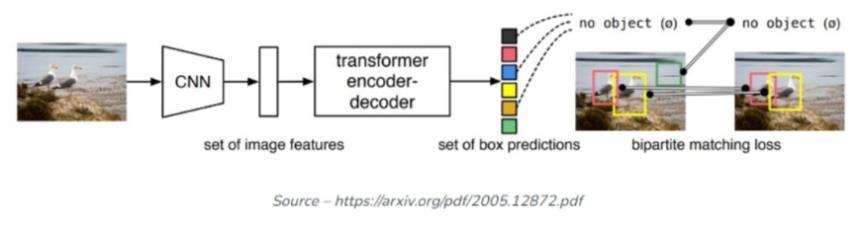}}
    \caption{DETR concept for image detection [20]}
    \label{fig:DETR_concept}
\end{figure}
\subsubsection{Image Detection with Deep Learning}
There is an abundant of research in the recent  past to detect objects in EO-IR imagery using deep learning techniques such as YOLO versions \cite{Singha_2021,Sahin_2021}, fast-Region-based Convolutional Neural Network (fast-RCNN) \cite{Nalamati_2019}, DEtection TRansformer (DETR), etc...\cite{Taha_Access20,Samaras_20,Samadzadegan_2022}.  As stated in the Introduction section, most of the deep learning techniques struggle to detect small drones when they are far away from the camera and occupy only a small number of pixels in the image.    Further, fine classification (e.g., multiple DJI drones) is also challenging except getting   a generic identifier such as 'drone'. Among the existing deep learning based image detectors, DETR  \cite{Carion_2020} is shown to have a good  accuracy and run-time performance trade-off  over other exiting deep learning models against the  challenging Common Objects in Context (COCO) object detection dataset.    Thus, we use DETR as the deep learning based image detector in this work.

\begin{figure}
    \centering
   {\includegraphics[width=0.4\textwidth]{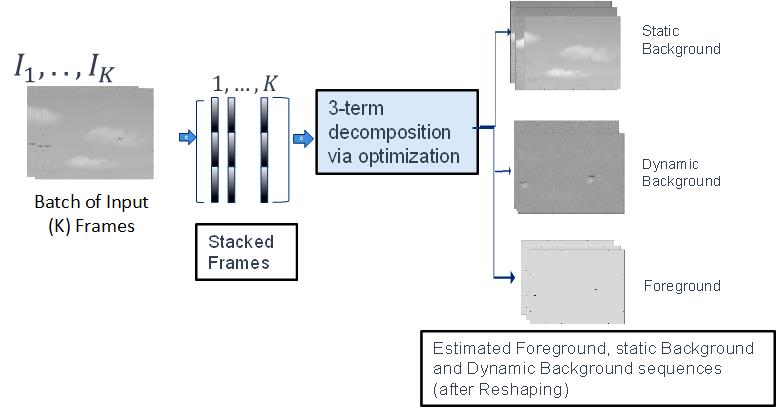}}
    \caption{Overview of RPCA for foreground/background separation with EO imagery }
    \label{fig:RPCA_Overview}
\end{figure}
\begin{figure}
    \centering
   {\includegraphics[width=0.4\textwidth]{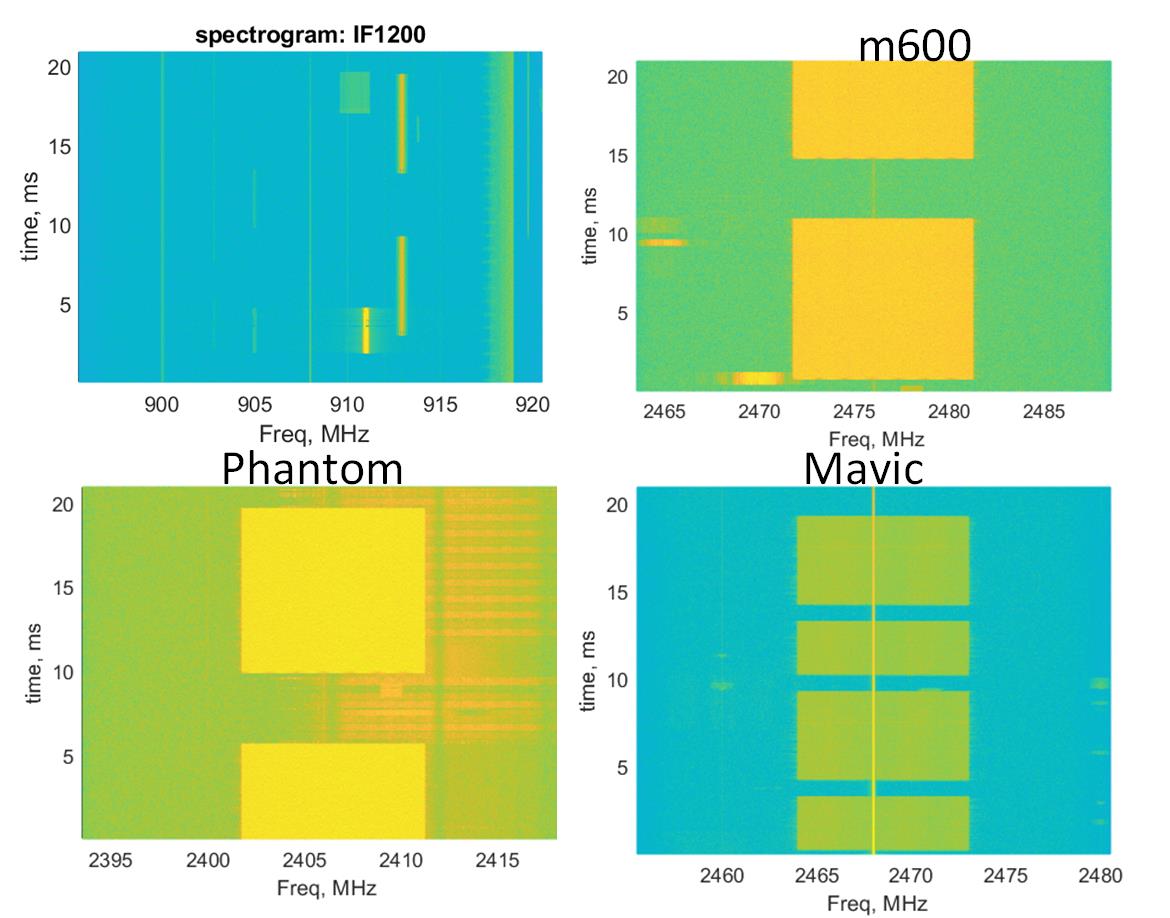}}
    \caption{Spectrogram of the 4 drones (from left to right; IF1200, m600, Phantom, Mavic),  Center frequencies of IF1200, m600, Phantom and Mavic are 908MHz, 2476MHz, 2406MH and 2468MHz, respectively}
    \label{fig:Drone_spectro}
\end{figure}
DETR is a set-based object detector using a Transformer on top of a convolutional backbone (Fig. \ref{fig:DETR_concept}). It uses a conventional CNN backbone to learn a 2D representation of an input image. The model flattens it and supplements it with a positional encoding before passing it into a transformer encoder \cite{Carion_2020}. The DETR model is loaded from huggingface.co, an online resource for sharing open-source models. Our DETR model was trained using pytorch lightning,
a machine learning (ML) framework built on top of pytorch optimized for multi-graphics processing unit (GPU) training. This allowed us to train models and iterate faster. Our training data is a drone object detection dataset from \cite{Pawelczyk_20}.
 This allowed us to train and iterate the model much faster than native pytorch, and improve the performance using various built-in tools to aid training such as learning rate schedulers and stochastic weight averaging \cite{Izmailov_2018}. The augmentations used were horizontal flip ($50\%$ of samples), color jitter ($25\%$), random brightness/contrast ($20\%$) and motion blur ($10\%$).

\subsubsection{Image Detection via Foreground/Background Separation}
There is a long history for detecting  moving objects in imagery using foreground/background separation techniques \cite{Piccardi_2004,Kulchandani_2015}. Robust Principal Component Analysis (RPCA) is a subspace method which is shown to perform better than other comparable techniques for  foreground detection with complex dynamic backgrounds  \cite{Guyon_2012,Oreifej_13,Candes_11}.  While RPCA is quite extensively investigated for image detection in video surveillance applications, there is no sufficient study in the literature to understand its capabilities in detecting drones, especially, when the drones are far away from the camera (in the region where most of the state-of-the-art deep learning techniques fail) and when the background is highly dynamic due to clouds, occlusions and the impact of other moving objects. In this work, we explore RPCA for drone detection with EO imagery. In general, RPCA   operates in batch mode where a collection of frames is processed at a time. Let $I_1, \cdots, I_K$ denote a set of $K$-frames captured by a camera with frame dimension $N_1\times N_2$. By stacking all the columns of a given frame in to a vector, a matrix $X\in R^{N\times K}$ is created where the $k$-th column of $X$ corresponds to the vectorized version of $I_k$ (Fig. \ref{fig:RPCA_Overview}) and $N=N_1 N_2$. In the RPCA framework, the  matrix $X$ is decomposed to a low-rank (background) matrix, a sparse (foreground) matrix and a dense error (dynamic background) matrix so that
\begin{figure}
\centering
\begin{subfigure}[b]{0.23\textwidth}
\centering
\includegraphics[width=\textwidth]{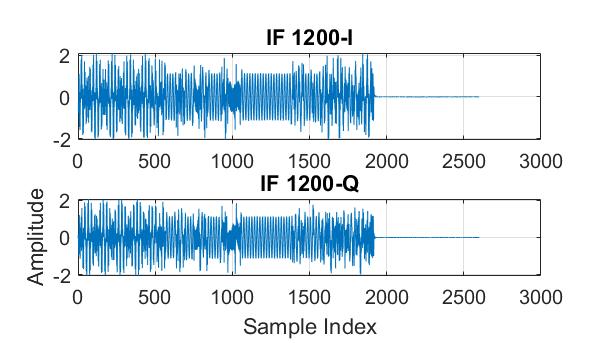}
\caption{IQ: IF 1200}
\label{Fig_DETR_r16}
\end{subfigure}
\hfill
\begin{subfigure}[b]{0.23\textwidth}
\centering
\includegraphics[width=\textwidth]{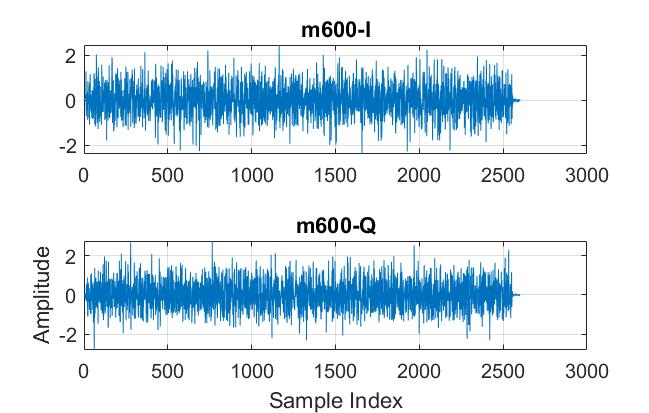}
\caption{IQ: DJI m600}
\label{Fig_DETR_r16}
\end{subfigure}
\hfill
\begin{subfigure}[b]{0.23\textwidth}
\centering
\includegraphics[width=\textwidth]{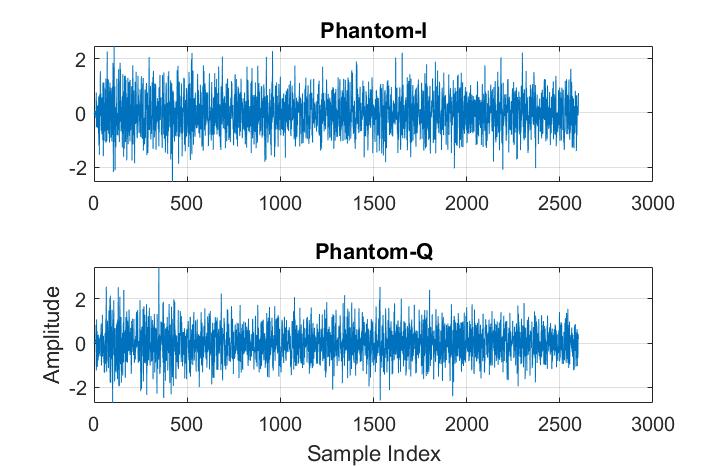}
\caption{IQ: DJI Phantom}
\label{Fig_DETR_r16}
\end{subfigure}
\hfill
\begin{subfigure}[b]{0.23\textwidth}
\centering
\includegraphics[width=\textwidth]{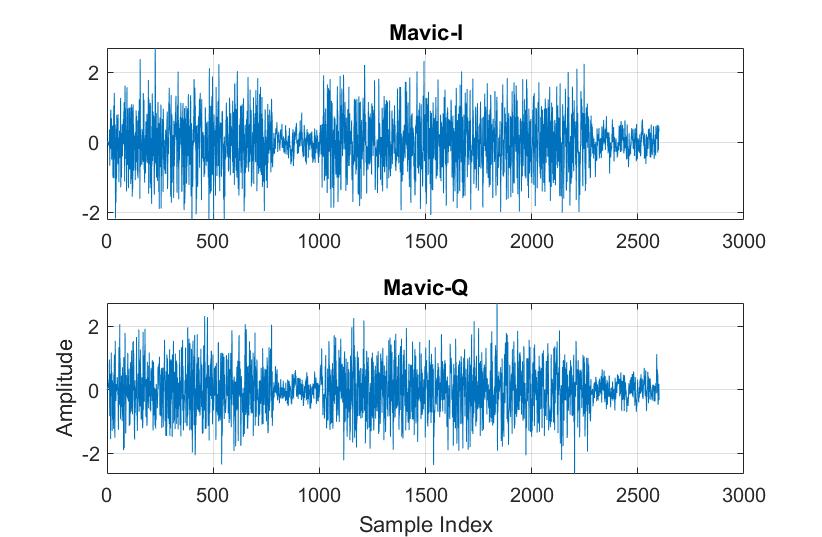}
\caption{IQ: DJI Mavic}
\label{Fig_RPCA_P_r16}
\end{subfigure}
\caption{Baseband IQ representation of the downlink waveforms of 4 drones }\label{fig_IQ}
\end{figure}

%\begin{figure}
%    \centering
%   {\includegraphics[width=0.48\textwidth]{figures/Range_vs_time_MAvic_Phantom.jpg}}
%    \caption{Range over time; r14}
%    \label{fig:r14_range_time}
%\end{figure}
\begin{eqnarray*}
  X=L+S+E
\end{eqnarray*}
where $L,S,E\in R^{N\times K}$ represent the low-rank, sparse and error matrix, respectively.   The three matrices $L,S,E$ are recovered by solving the  following optimization problem:
\begin{eqnarray}
  \min_{L,S,E} ||L||_{\star} + \tau ||S||_1 + \lambda ||E||_F^2 \nonumber\\
  \mathrm{such}~  \mathrm{that} ~  X=L+S+E \label{eq:RPCA_opt}
\end{eqnarray}
where $||\cdot||_{\star}$,  $||\cdot||_{1}$ and  $||\cdot||_{F}$   denote the nuclear  norm, $l_1$ norm and the Frobenius norm  of a matrix, respectively, and $\tau$ and $\lambda$ are penalty parameters.  Finding the global optimal solution of (\ref{eq:RPCA_opt}) in closed-form is a difficult problem, and alternating-direction-method of multiplier (ADMM) is a widely used approach which solves (\ref{eq:RPCA_opt}) iteratively by optimizing one variable keeping the rest constant \cite{Oreifej_13} (See Appendix).
The ADMM based foreground/background separation algorithmic steps are summarized in Algorithm \ref{Alg_ADMM}.

\begin{algorithm}
\caption{Foreground/Background Separation via ADMM}
Inputs; $X$,  scaler ~parameters~ $\lambda$, $\tau$, $\rho$ \\
Initialization: $\mathbf Y$, $S$, $E$,$\beta > 0$,\\

\textbf{While} ~not ~converged ~ \textbf{do}\\

$UWV^T = svd\left(X-E^k - S^k +\frac{Y^k}{\beta^k}\right)$\\
$L^{k+1} = U\Pi_{\frac{1}{\beta}}(W) V^T$\\
$S^{k+1} = \Pi_{\frac{\tau}{\beta^k}}\left(X-E^k+\frac{Y^k}{\beta_1^k}- L^{k+1}\right)$\\
$E^{k+1}=(1+2\lambda/\beta^k)^{-1} (X - L^{k+1} - S^{k+1}+Y^k/\beta^k)$\\

$Y^{k+1}=Y^k + \beta^k (X-L^{k+1} - S^{k+1} - E^{k+1})$\\

$\beta^{k+1}=\rho\beta^k$\\
\textbf{End}\\
$\Pi_{\alpha}(x) = sign(x).max\{|x|-\alpha,0\}$\\
$svd$: singular value decomposition \\
Outputs: $\hat{S}, \hat{L}, \hat{E} $\\
 \label{Alg_ADMM}
\end{algorithm}

While  RPCA is a promising approach for foreground and background separation, its computational complexity is quite significant (due to the  need of singular value decomposition) compared to other basic techniques for foreground/background separation  when the image frame dimension is large. In order to implement RPCA in real-time, we run RPCA after partitioning the original images where individual partitions can be processed in parallel. After processing each batch with parallel partitions, the individual partitions  are mapped back to the original image dimension, so that RPCA can be implemented in near real time.
%\begin{figure}
%    \centering
%   {\includegraphics[width=0.48\textwidth]{figures/Sensor_layout.jpg}}
%    \caption{Sensor layout for scenarios r14 and r16}
%    \label{fig:r14_Sensor_layout}
%\end{figure}
\subsection{Drone Detection, Localization and Identification  with Passive RF}
Most  drones (unless they full autonomous)  use a downlink communication protocol to relay messages to a remote controller. While drones may not transmit continuously depending on the intent of the drone, the purpose is to leverage the useful information inferred from passive RF  whenever available to support the  tracking and identification capabilities with only EO data.  As shown in Fig. \ref{fig:Drone_spec}, the 4 drones considered in this work use proprietary waveforms for downlink communication.

\subsubsection{Drone Identification via RF Fingerprinting}
In Fig. \ref{fig:Drone_spectro}, the spectrogram is shown for the 4 drones considered. It is noted that DJI m600 and Phantom share very similar timing and frequency characteristics, while DJI Mavic and IF 1200 are quite different from the first two. Nevertheless, we train a deep learning model to classify these 4 drones extracting the RF fingerprint from the baseband IQ-level data. The  deep learning model  operates with raw IQ data and pre-processing as discussed below is done to extract IQ vectors for training/test.
%\begin{figure}
%    \centering
%   {\includegraphics[width=0.48\textwidth]{figures/r14_sensor_tgt_layout.jpg}}
%    \caption{Sensor and target locations  wrt to EO sensor; cyan circles: PRF payload sensors (d104-d106), green circles; PRF tower sensors (d101-d103), red star: EO sensor (d004), red line-DJI Phantom track, blue line-DJI Mavic track}
%    \label{fig:r14_sensor_target_wrtEO}
%\end{figure}

\subsubsection{Localization}
With  multiple passive RF sensors, we explore a TDOA-based approach for drone localization.
In order to get 3D position estimates  of the drones, at least 4 passive receivers are necessary.

\subsubsection*{Passive RF Data pre-Processing for RF Fingerprinting and Localization}
The data is captured at $25MHz$ centered at center frequencies  listed in Fig. \ref{fig:Drone_spectro} for each drone.
It is noted that the data link used by DJI drones are pretty wideband  while the waveform used by IF 1200 has a hop pattern within the $25MHz$ bandwidth. The next step is to identify the center frequency of the hop (mainly for the IF1200, since for DJI drones, it is fixed), and extract a baseband IQ vector of  $250kHz$ wide  by low pass filtering  and downsampling the data at Nyquist rate. The reason for selecting a relatively small frequency band for RF fingerprinting is to  get a relatively a moderate size IQ vector over  a reasonable time interval to capture rising and falling edges of the transmitted waveforms.  By observing the packet length of all four drones, the decision making time for RF fingerprinting is taken as $0.0210s$ which corresponds to $5250$ samples at $250 kHz$ sampling rate. After edge detection of the extracted  IQ vector, the noise component is removed so that the effective length of IQ is set to $2600$.   The pre-processed waveforms   used for training/test  are shown in Fig. \ref{fig_IQ}.

For TDOA computation using passive RF IQ, the original data is down converted to  $10MHz$ and perform noise filtering to extract only the desired signal component at every $0.0210s$.  TDOAs are estimated by computing the cross correlation of each sensor pair  considered.

\subsection{Detection-Detection Association and Tracking with  EO and Passive RF in 2D}
Let $(X,Y,Z)$ be the world coordinates of the target  with respect to the camera position  and $(x,y)$ be the corresponding location in the image pixel plane. Then, $(x,y)$ can be expressed as
\begin{eqnarray}
w\times [x~ y~ 1] = [X~Y~Z~1] \times [R; T]\times K \label{eq_camMatrix}
\end{eqnarray}
where $K$ is the camera intrinsic matrix, $R$ is the camera rotation matrix, $T$ is a translation  vector, $w$ is a scaler. In ESCAPE 2021 data collect, the camera intrinsic matrix parameters are recorded and calibration is done to estimate the other extrinsic parameters ($R$ and $T$).

The estimated target locations with respect to the EO sensor are  projected onto the 2D image plane using (\ref{eq_camMatrix}). Then,  the passive RF detections and EO detection are ordered in time in the 2D image plane and,  a 2D Kalman filter with a constant velocity model is used for 2D tracking.

\section{Performance Analysis}\label{sec_performance}
This section provides an overview of the experiments and overall performance of individual modalities as well as with the fusion architecture. Experimentation  consist of multiple steps; (i) train a deep learning model to obtain drone fingerprint using passive RF IQ data, (ii) TDOA estimation and localization with passive RF IQ (iii). 2D detections via  RPCA and DETR with EO imagery (iv) End-to-End architecture testing.

To train the deep learning classifier for RF fingerprinting, we consider the scenarios listed in Table \ref{Table_scene_FP}. It is noted that, we follow a similar  scenario and device  labeling as in \cite{Zulch_23} just for simplicity.
\begin{table}
\renewcommand{\arraystretch}{1.3}
\caption{\bf Scenarios  and Targets used to Train Deep Learning Model for RF Fingerprinting }
\label{RF_FP_data}
\centering
\begin{tabular}{|c|c|c|c|}
\hline
\bfseries Scenario  & \bfseries Target  & Devices  & passes \\
\hline\hline
r15              & IF 1200 & 101, 102, 103 & 01, 02\\
r15               & m600 & 104, 105, 106 & 01, 02\\
\hline
r13              & Mavic & 101, 102, 103 & 01, 02\\
r13               & Phantom & 104, 105, 106 & 01, 02\\
\hline
\end{tabular}\label{Table_scene_FP}
\end{table}

 Each collect per target has multiple passes to impose redundancy which is useful mainly in deep learning exercises. Each scenario is  $~50s$ long, which provides $\sim 2300$ IQ examples on average  with $0.0210s$ decision making time for RF fingerprinting (some scenarios can be longer than this). We consider different devices and different passes to ensure diversity across communication channels. With the scenario selection as in Table \ref{Table_scene_FP}, each target has $\sim 11,000$ labeled IQ vectors for training.
\begin{table}
\renewcommand{\arraystretch}{1.3}
\caption{\bf Scenario  and Target  Description used to Test End-to-End EO-RF Fusion Architecture}
\label{DueDates}
\centering
\begin{tabular}{|c|c|c|c|}
\hline
\bfseries Scenario  & \bfseries Targets & Flight path & Distance \\
\hline\hline
r06p01               & Phantom & North to South & $\sim 500m$ \\
r14p01               & Phantom, Mavic & North to South & $\sim 500m$ \\
r16p01 & m-600, IF 1200 & North to South & $\sim 500m$\\
\hline
\end{tabular}\label{Table_End_End}
\end{table}

\begin{figure}
\centering
\begin{subfigure}[b]{0.4\textwidth}
\centering
\includegraphics[width=\textwidth]{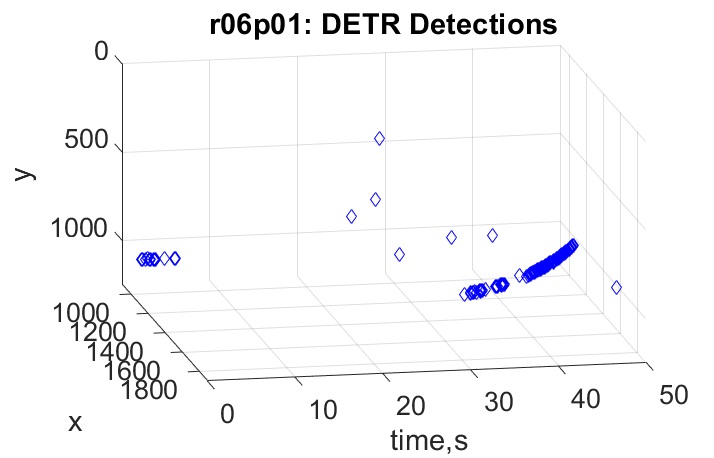}
\caption{r06; DETR 2D Detections}
\label{Fig_DETR_r06}
\end{subfigure}
\hfill
\begin{subfigure}[b]{0.40\textwidth}
\centering
\includegraphics[width=\textwidth]{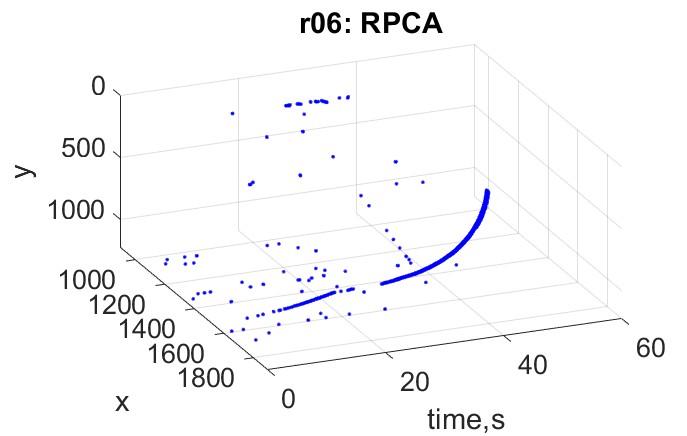}
\caption{r06; RPCA 2D Detections}
\label{Fig_RPCA_P_r06}
\end{subfigure}
    \caption{r06, Estimated foreground mask over time (range) with different algorithms}
    \label{fig:r06_2D_Detections}
\end{figure}

\begin{figure}
    \centering
   {\includegraphics[width=0.40\textwidth]{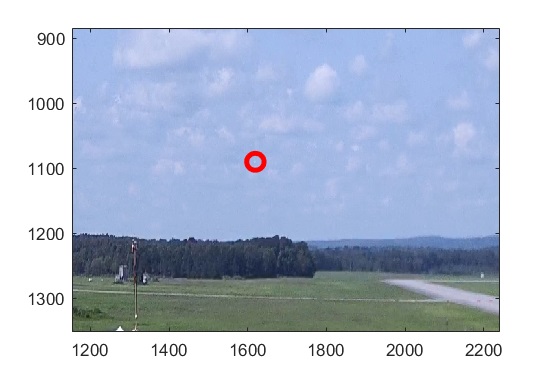}}
    \caption{First Detect of Phantom in r04 with RPCA; range $~435m$ from EO004 camera.}
    \label{fig:r06_firstDetect}
\end{figure}
\begin{figure}
\centering
\includegraphics[width=0.40\textwidth]{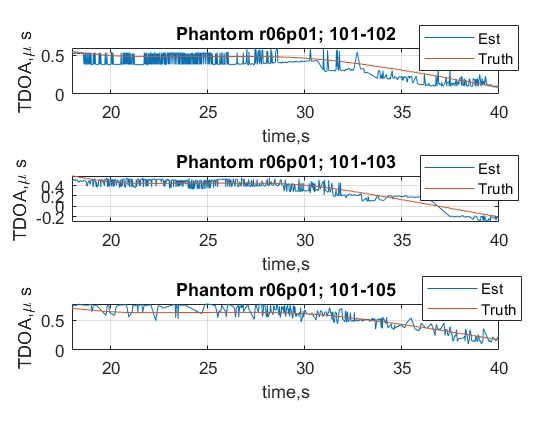}
\caption{r06; TDOA estimates with sensor pairs d101-d102, d101-d103, d101-d105; Time is  with respect to the start time of the EO camera E001 recording; for r06, there is a time lag between EO and passive RF recordings}
\label{Fig: TDOA_r06p01}
\end{figure}

\begin{figure}
\centering
\includegraphics[width=0.48\textwidth]{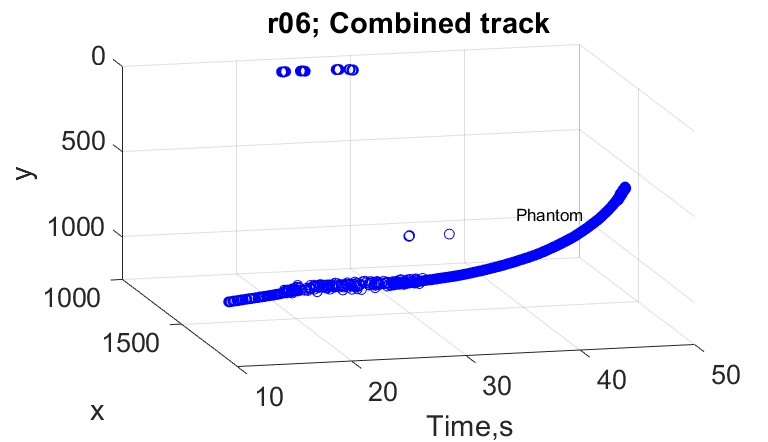}
\caption{r06; Combined track with associating passive RF projections with 2D Image detections}
\label{Fig:combine2D_track_r06p01}
\end{figure}

To test the end-to-end architecture for detection localization and tracking, three  scenarios are considered as depicted in Table \ref{Table_End_End}. In scenario 1 (r06p01), only DJI Phantom flies from North to South. In Scenario 2 (r14p01), DJI Mavic and DJI Phantom drones fly at a constant altitude over $\sim500m$ while in Scenario 2 (r16p01), IF 1200 and m600 are flying along the same path as in r14p01.

\subsection{Analyzing r06}
\subsubsection{r06: Drone Detection Performance  with EO Imagery}
In Scenario r06, Phantom starts flying  from north at a distance of $\sim 500m$ from the  EO sensor which is located at south. Fig. \ref{fig:r06_2D_Detections} illustrates the RPCA and DETR  detections over 2D image plane over time (and range). The EO data is sampled at $30Hz$ and has a frame size of $3840\times 2160$ (full frame is not shown in the figure for clarity).  It is noted that detections with r06 EO imagery is quite challenging due to the slow moving clouds. Fig. \ref{fig:r06_firstDetect} shows the original image when the RPCA first detects the target where the drone is $~435m$ away from the camera. Note that RPCA parameters are optimized to obtain reliable detections over the entire trajectory, so some clutter detections (especially coming from the coulds) are visible towards the beginning  of the trajectory, which go away with Kalman Filtering as shown later in this section. DETR struggles to get reliable detections until the drone is about 200m away from the camera.

\subsubsection{r06: Drone Classification Performance with RF Fingerprinting} Classification performance of Phantom using the trained model to classify 4 drones using passive RF IQ similar to the scenario  discussed under next scenario (Analyzing r14), thus, is avoided here for brevity.
\subsubsection{r06: TDOA Based 3D Localization}
In r06 where a single drone is flying, all six passive RF receivers are tuned to the same downlink frequency  of the drone, $2.406
GHz$. For 3D localization, the sensors 101, 102, 103, and 105 are used. With $3$ independent pairs of TDOA,  spherical intersection algorithm \cite{Schau_87} is used  to find the 3D drone  position. Fig. \ref{Fig: TDOA_r06p01} illustrates the TDOA estimates for ro06 (compared to the ground truth compared with actual distances) while Fig. \ref{Fig:combine2D_track_r06p01} illustrates the combined track in the 2D plane. It is noted that some of discontinuities observed in EO only detection is compensated for after integrating passive RF and the true track is given a unique ID. It is worth noting that, since passive RF has a time lag compared to the EO recording in this scenario, early part of the trajectory is still missed.
\begin{figure}
    \centering
   {\includegraphics[width=0.4\textwidth]{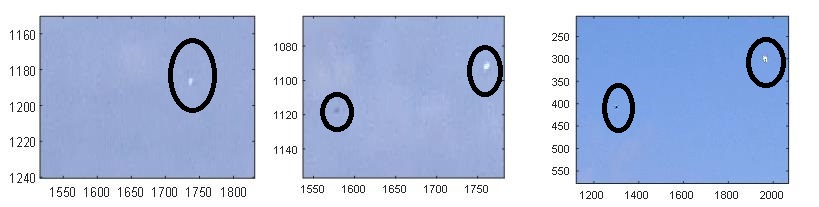}}
    \caption{r14: DJI Mavic (negative contrast) and DJI Phantom (Positive contrast) in the same image frame; different range from the EO camera location; (Left) at $\sim 450m$, (middle) $\sim 300m$, (right) at $\sim 200m$.  It is noted that, Mavic is not captured at all in the  image plane at range $\sim 450m$.}
    \label{fig:r14_zoomed_image}
\end{figure}
\begin{figure}
\centering
\begin{subfigure}[b]{0.40\textwidth}
\centering
\includegraphics[width=\textwidth]{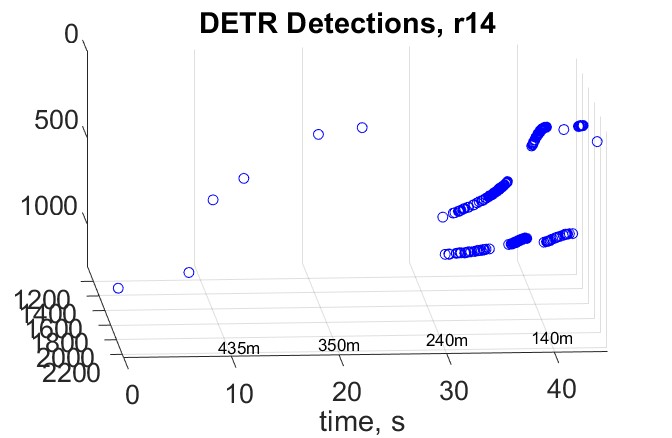}
\caption{ DETR, both Mavic and Phantom are reliably detected around $\sim 240m$ away from the camera}
\label{Fig_DETR_r16}
\end{subfigure}
\hfill
\begin{subfigure}[b]{0.40\textwidth}
\centering
\includegraphics[width=\textwidth]{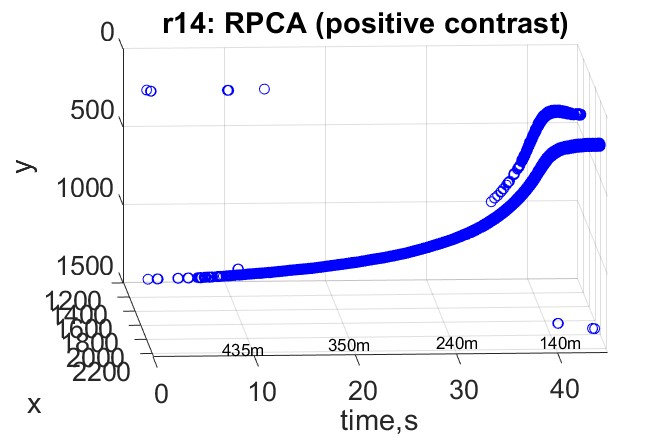}
\caption{ RPCA, positive contrast, Phantom is detected at $\sim 440m$, while Mavic is not detected early}
\label{Fig_RPCA_P_r16}
\end{subfigure}
\hfill
\begin{subfigure}[b]{0.40\textwidth}
\centering
\includegraphics[width=\textwidth]{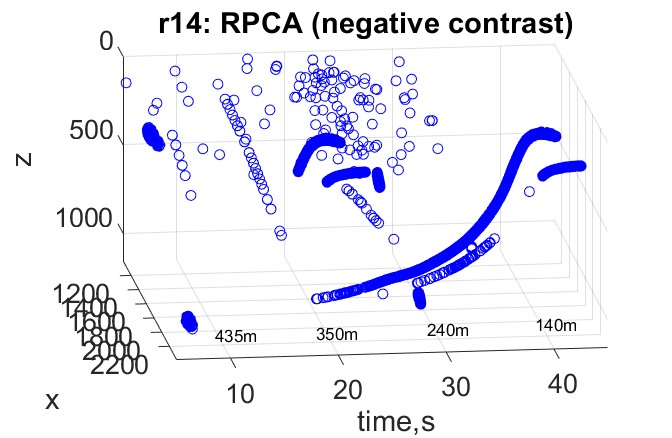}
\caption{RPCA, negative contrast, Mavic is first detected at $\sim 350m$, while Phantom is not detected early}
\end{subfigure}
\caption{r14, Estimated foreground mask over time (range) with different 2D image detection algorithms}\label{Fig_RPCA_N_rr14}
\end{figure}
\begin{figure}
    \centering
   {\includegraphics[width=0.48\textwidth]{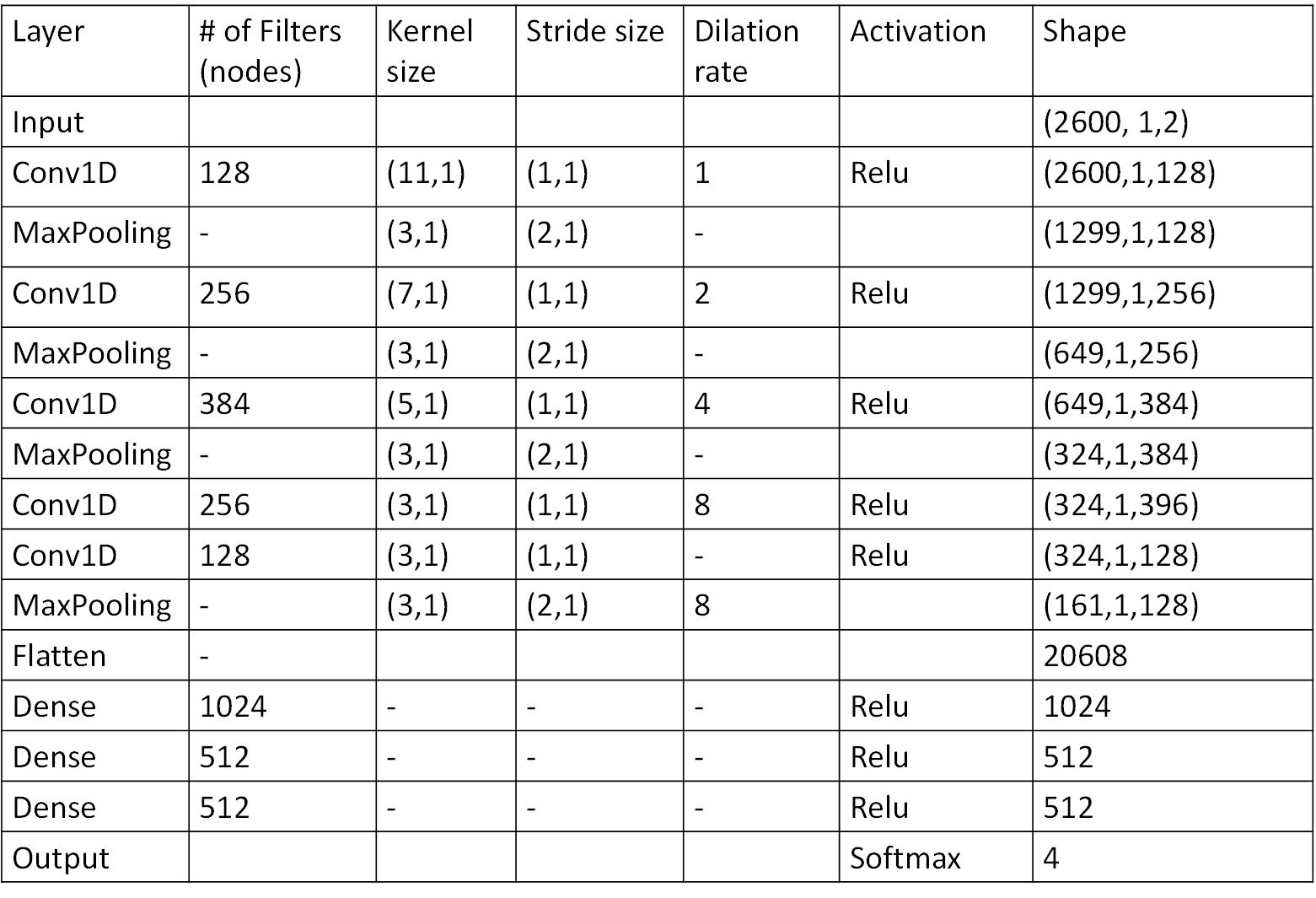}}
    \caption{Configuration of the developed Deep Learning model for RF fingerprinting}
    \label{fig:DL_arch}
\end{figure}

\subsection{Analyzing r14}
%For r14, the actual sensor locations and the ground truth target with respect to the EO camera (reference point) for scenario r14 are illustrated in Fig. \ref{fig:r14_sensor_target_wrtEO}.
%In r14, two drones,  Phantom and Mavic are flying.
In r14, Mavic and Phantom drones fly simultaneously starting at a range of $\sim 500m$ away from the EO sensor.
%In Fig. \ref{fig:r14_range_time}, the range over time is shown.
%\begin{figure}
%\centering
%\begin{subfigure}[b]{0.45\textwidth}
%\centering
%\includegraphics[width=\textwidth]{figures/Model_conf_test_Mavic_2.png}
%\caption{Testing Mavic at Device 101}
%\label{Fig_DETR_r16}
%\end{subfigure}
%\hfill
%\begin{subfigure}[b]{0.45\textwidth}
%\centering
%\includegraphics[width=\textwidth]{figures/Model_conf_test_Phantom_2.png}
%\caption{Testing Phantom at Device d104}
%\label{Fig_RPCA_P_r16}
%\end{subfigure}
%    \caption{Confidence of the trained Deep Learning model for RF fingerprinting  for each trained class when tested with (a) Mavic (b) Phantom in r14p01}
%    \label{fig:r14_RF_FP_model_confidence}
%\end{figure}
\subsubsection{r14: Drone Detection Performance  with EO Imagery}
It is noted that Mavic and Phantom drones are light weight (Fig. \ref{fig:Drone_spec}) and very small which are very hard to detect when they are far away from the camera. Zoomed images at different ranges from the camera  are shown in Fig.  \ref{fig:r14_zoomed_image} to depict the point-target nature of Mavic and Phantom  at various ranges.

As seen in Fig. \ref{fig:r14_zoomed_image}, the two drones have positive and  negative contrast with respect to  the background image, thus it is hard to optimize the image detection algorithms to detect both simultaneously. For RPCA, we consider this observation as a side information and optimize the algorithm to detect objects with positive contrast and the object with negative contrast in parallel  and fuse them together. For RPCA, we used 30 frames at once for batch processing (i.e. $1s$ of latency with $30Hz$ frame rate). Fig. \ref{Fig_RPCA_N_rr14} shows the estimated foreground mask (binary value indicating the centroid of the bounding box) over time (range) for scenario r14p01.

\begin{table*}
\renewcommand{\arraystretch}{1.3}
\caption{\bf Average Confidence of the Trained Deep Learning Model for RF Fingerprinting  when Tested with Mavic and Phantom in r14p01}
\label{DueDates}
\centering
\begin{tabular}{|c|c|c|c|c|}
\hline
\bfseries Tested  & \bfseries Avg Conf  &\bfseries Avg Conf &\bfseries Avg Conf  &\bfseries Avg Conf  \\
\bfseries Class  & \bfseries on IF 1200 &\bfseries on Mavic&\bfseries on Phantom &\bfseries on m600 \\
\hline\hline
Mavic               & 0.002 & 0.988&0.011 & 0 \\
\hline
Phantom  & 0.007 & 0.004 & 0.995  & 0.001\\
\hline
\end{tabular}\label{Table_Avg_Conf_r14}
\end{table*}
From Fig. \ref{Fig_RPCA_N_rr14}, it is observed that the DETR algorithm fails to detect the drones until they    are about $240m$ away from the camera. The RPCA algorithm on the other hand, can detect both drones much earlier (Mavic at $\sim 350m$ with negative contrast and Phantom at $\sim 440m$ with positive contrast). Note that what is shown in Fig. \ref{Fig_RPCA_N_rr14} are raw detections in the image plane,  and  some of the isolated false alarms could be eliminated once the tracker is run. This is done with the composite detections after integrating with  passive RF below; however, in order to get EO-only tracks, a 2D Kalman filter can run  on the raw EO detections.

\subsection{r14: Drone Classification Performance with RF Fingerprinting}
This section discusses the device ID capability of passive RF for scenario $r14$.  The developed  deep learning model consists of a set of convolutional, pooling and dilation layers and the architecture and parameters are summarized  in Fig. \ref{fig:DL_arch}. As discussed in Section \ref{sec_overview}, the designed deep learning based classifier is trained with all 4 drone  classes in this ESCAPE data collect using the scenarios listed in Table \ref{Table_scene_FP}. It is noted that the training data is constructed using $r13$ and $r15$ where $r13$ contains the same targets as in $r14$ while  $r15$ contains the same targets as in $r16$. However, in $r13$ and $r15$, drones fly south to north while in $r14$ and $r16$, the drones fly north to south. For the test phase, we consider the scenarios $r14$ and $r16$.

In the test phase, the trained model gives a confidence value between $[0,1]$   for each trained class that the model believes the test vector belongs. The average confidence for each trained class (over all the time indices) in the trajectory when tested with Mavic and Phantom is summarized in Table \ref{Table_Avg_Conf_r14}. It is worth noting that  the output layer of the trained model gives a confidence value which can be thresholded (or take the maximum over all trained classes) to declare the corresponding drone class.  It can be seen that while few examples of Mavic get confused with Phantom, overall, the trained model provides a promising confidence for the true class.

%In Fig. \ref{fig:r14_RF_FP_model_confidence}, the model confidence for each trained class when tested with Mavic and Phantom IQ vectors in scenario r14p01 is illustrated over the time index. The x-axis is the example index over time where each example is computed at every  $0.0210s$. It is noted that the results in Fig. \ref{fig:r14_RF_FP_model_confidence} are based on the passive RF receivers d101 and  d104 for Mavic and Phantom, respectively, which are the farthest passive RF receivers from the drone start location.
%
%From Fig \ref{fig:r14_RF_FP_model_confidence}, it can be seen that while few examples of Mavic get confused with Phantom, overall, the trained model provides a promising confidence for the true class.
%\subsection{TDOA based localization in 3D Sensor Plane}
%
%\subsection{Projection to the 2D Sensor Plane}

\subsection{r14: Final Tracker in the 2D Plane with Passive RF and EO Fusion }
As observed in Fig. \ref{Fig_RPCA_N_rr14},  while RPCA based EO detections perform much better than DETR, it still struggles to detect Mavic at early stages in r14. Also, RPCA does not do any classification, but only detection.
To incorporate the device ID obtained via RF fingerprinting and compensate for the poor performance at the far range in detection with only EO, we performed TDOA based localization until the drones are about $~300m$ away from the EO camera (i.e. about $~30s$ after drones start flying).
In Scenario r14, only 3 passive RF receivers are tuned to one drone, thus the number of sensors is not sufficient to perform 3D localization with spherical intersection.  It is noted that the drones are flown at a constant altitude in this data collect, thus, using that as a side information, three passive RF receivers are used to estimate the other two dimensions of the targets  in the sensor plane via maximum likelihood estimation.
After TDOAs are estimated using passive RF IQ (by computing the cross correlation among sensors), the locations are found using the maximum likelihood approach. Let $\Delta \tau_i$ be the TDOA of the $i$-th sensor pair. Then, the location of the target with respect to a reference point is estimated as
\begin{eqnarray*}
\hat{r}_t = \underset{r_t}{arg min} \left\{\sum_{i=1}^L\left(\frac{||r_t - r_{i,1}||_2 - ||r_t -r_{i,2}||_2}{c} - \Delta \tau_i\right)^2  \right\}
\end{eqnarray*}
where $L$ is the number of sensor pairs, $c$ is the speed of light, $r_t=[X_t, Y_t, Z_t]$ is the location of the target and the $r_{i,1}$ is the location of the $1st$ sensor in the $i$-th sensor pair, with respect to a reference point.

\begin{figure}
    \centering
   {\includegraphics[width=0.4\textwidth]{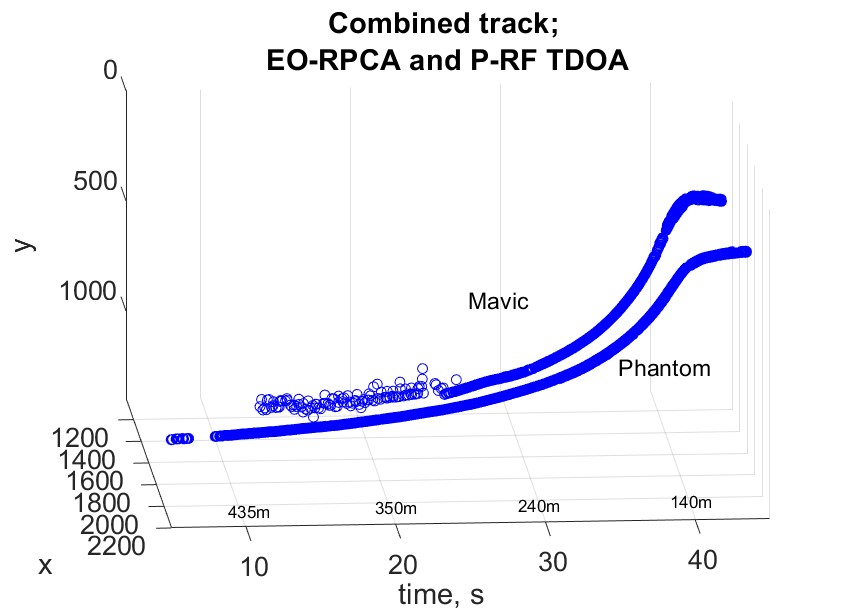}}
    \caption{r14: Final track with EO and P-RF fusion, passive RF location estimates are used until the EO only forms a consistent track for each drone, device ID is associated via RF fingerprinting}
    \label{fig:r14_combined_track}
\end{figure}

For Mavic, the passive RF sensors d101-d103 are  used while for Phantom devices d104-d106 are used.
%\begin{figure*}
%\centering
%\begin{subfigure}[b]{0.45\textwidth}
%\centering
%\includegraphics[width=\textwidth]{figures/TDOA_truth_vs_est_mus_Mavic_r14_p01.jpg}
%\caption{Mavic}
%\label{Fig_DETR_r16}
%\end{subfigure}
%\hfill
%\begin{subfigure}[b]{0.45\textwidth}
%\centering
%\includegraphics[width=\textwidth]{figures/Model_conf_test_Phantom.png}
%\caption{Phantom}
%\label{Fig_RPCA_P_r16}
%\end{subfigure}
%    \caption{Estimated TDOA for the sensor pairs considered (The time reference is with respect to the passive RF sensors)}
%    \label{fig:r14_RF_FP_model_confidence}
%\end{figure*}

%\begin{figure}
%\centering
%\begin{subfigure}[b]{0.45\textwidth}
%\centering
%\includegraphics[width=\textwidth]{figures/FG_mask_detections_r16_detr_new.jpg}
%\caption{DETR}
%\label{Fig_DETR_r16}
%\end{subfigure}
%\hfill
%\begin{subfigure}[b]{0.45\textwidth}
%\centering
%\includegraphics[width=\textwidth]{figures/r16_RPCA_detections_new3_pos.jpg}
%\caption{RPCA, positive contrast}
%\label{Fig_RPCA_P_r16}
%\end{subfigure}
%\hfill
%\begin{subfigure}[b]{0.45\textwidth}
%\centering
%\includegraphics[width=\textwidth]{figures/r16_RPCA_detections_new3_neg.jpg}
%\caption{RPCA, negative contrast}
%\label{Fig_RPCA_N_r16}
%\end{subfigure}
%\caption{r16, Estimated foreground mask with EO over time  with different algorithms}
%\label{Fig_EO_Det_N_r16}
%\end{figure}
It is worth noting  that the P-RF sensor recordings have a lag ($\sim 12s$) compared  to the EO recordings, thus passive RF detections are shifted by $12s$ to be synchronised with EO detections. Also, for the devices d104-d106, there is a time offset about $6s$ between $d106$ and the other two which was appropriately shifted. It is further worth noting that, in ESCAPE 2021 data collect, only three passive RF sensors are monitoring any given drone  when two drones are flying simultaneously. Ideally, at least 4 sensors are needed to do 3D geolocation using TDOA. However, using the fact that the all drones in ESCAPE 2021 data collect fly at approximately a constant altitude, we used three passive RF receivers to do 3D localization using the altitude  as a side information (with some uncertainty). Taking all these factors into account, and after performing detection-detection association in the 2D image plane, the final composite tracks for r14 are obtained as shown in Fig. \ref{fig:r14_combined_track}. Regarding track labeling, until  passive RF is integrated, the EO detections are associated via the Hungarian algorithm \cite{Burkard_2009} to associate detections over frames which provides a notional track ID for multiple tracks. Once  passive RF is integrated and associated with EO detections, the corresponding track labels are replaced by the passive RF device labels obtained via fingerprinting. In case where there are EO detections available before passive RF  detections  (due to offsets of the two types of sensors and/or  when drones are silent), tracks are labeled only with a notional ID obtained via the Hungarian data association with EO.
\begin{table*}
\renewcommand{\arraystretch}{1.3}
\caption{\bf Average Confidence of Trained Deep Learning Model for RF Fingerprinting  when Tested with IF1200 and m600 in r16p01}
\label{DueDates}
\centering
\begin{tabular}{|c|c|c|c|c|}
\hline
\bfseries Tested  & \bfseries Avg Conf  &\bfseries Avg Conf &\bfseries Avg Conf  &\bfseries Avg Conf  \\
\bfseries Class  & \bfseries on IF 1200 &\bfseries on Mavic&\bfseries on Phantom &\bfseries on m600 \\
\hline\hline
IF1200               &1 &0 &0. & 0 \\
\hline
m600  & 0 & 0 & 0.0002  & 0.9998\\
\hline
\end{tabular}\label{Table_Avg_Conf_r16}
\end{table*}

%\begin{figure}
%\centering
%\begin{subfigure}[b]{0.45\textwidth}
%\centering
%\includegraphics[width=\textwidth]{figures/Model_conf_test_IF1200_2.png}
%\caption{Testing IF1200 at Device 101}
%\label{Fig_DETR_r16}
%\end{subfigure}
%\hfill
%\begin{subfigure}[b]{0.45\textwidth}
%\centering
%\includegraphics[width=\textwidth]{figures/Model_conf_test_m600_2.png}
%\caption{Testing m600 at Device d104}
%\label{Fig_RPCA_P_r16}
%\end{subfigure}
%    \caption{Confidence of the trained Deep Learning model for RF fingerprinting  for each trained class when tested with (a) IF1200 (b) m600 in r16p01}
%    \label{fig:r16_RF_FP_model_confidence}
%\end{figure}
\begin{figure}
    \centering
   {\includegraphics[width=0.4\textwidth]{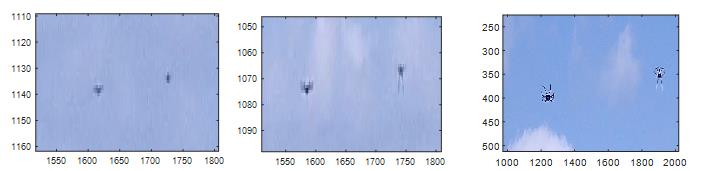}}
    \caption{r16: m-600 and IF1200 in the same image frame; different ranges from the EO camera location, (left) $\sim 450m$, (middle) $\sim 300m$, (right) $\sim 200m$}
    \label{fig:zoomed_r16}
\end{figure}

\subsubsection*{Analyzing r16}
In r16, IF1200 and m600 fly simultaneously. These two drones are relatively larger and heavier than the ones in r14. Zoomed images from  the EO sensor location  are shown in Fig. \ref{fig:zoomed_r16}.

%The EO only detections for r16 are shown in Fig. \ref{Fig_EO_Det_N_r16} with DETR and RPCA.

While detailed figures are omitted for brevity, RPCA when optimized to detect objects with  both positive and negative contrast with respect to the background is capable of detecting IF1200 and m600 few seconds after they appear in image data.   Compared to Mavic and Phantom considered in r14, IF1200 and m600 are relatively large and have both negative and positive contrast compared to the image background, thus, fairly well detected in both versions of RPCA. DETR still finds it difficult to detect both IF1200 and m600 at the early stages.

In order to assign a label to each detected drone in the EO image plane, RF fingerprinting results are associated with EO detections.
%In Fig. \ref{fig:r16_RF_FP_model_confidence}, the trained RF fingerprinting model confidence is shown over the time index when
When tested with IF1200 (d101) and m600 (d104) in r16,  Table \ref{Table_Avg_Conf_r16} summarizes the average confidence over time which again shows the promise of proposed RF fingerprinting  model for correct device identification.

\begin{figure}
    \centering
   {\includegraphics[width=0.4\textwidth]{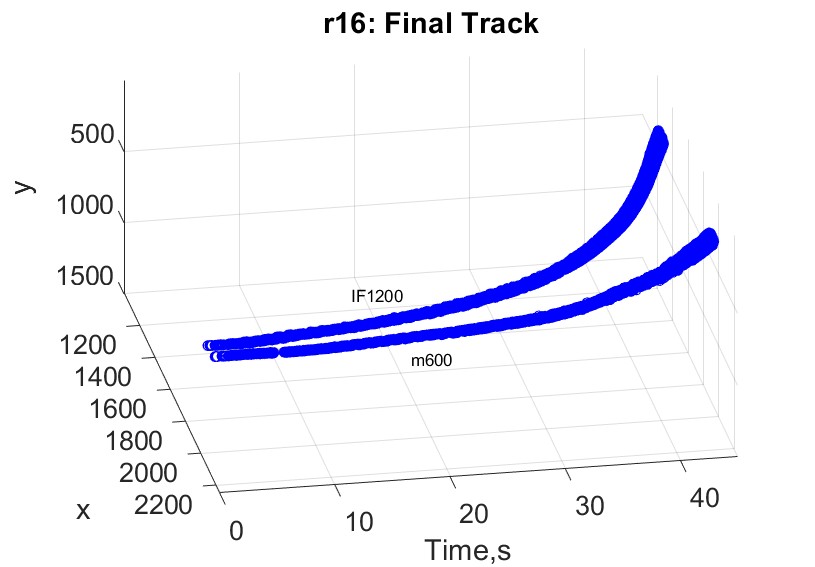}}
    \caption{r16: Final 2D track  with EO and P-RF fusion, passive RF location estimates are used until the EO only forms a consistent track for each drone, device ID is associated via RF fingerprinting}
    \label{fig:r16_combined_track}
\end{figure}

Since the EO can detect the drones in r16 at very early stages using RPCA, passive RF based localization is used only until the device ID is associated with the EO detections for both drones. The combined tracker after running a Kalman filter on the 2D detections is shown Fig. \ref{fig:r16_combined_track}.
\section{Discussion}
Based on the numerical studies done in this paper with the AFRL  ESCAPE-2021 multi-modal dataset, we make several observations regarding detection, identification  and tracking of Group 1 and 2 drones.
\begin{enumerate}
\item With only EO, Group 1 drones (DJI Phantom/Mavic) can be much better detected by advanced foreground/background separation algorithms compared to deep learning based techniques when the drones occupy only a small number of pixels in the image plane.  RPCA can struggle detecting very small drones (e.g. Mavic in the very early portion of the trajectory) which can be compensated if passive RF is available
    \item By incorporating device ID using passive RF, unique track ID could be obtained for the 2D track in the image plane. Thus, even when the passive RF is not available continuously, EO can still provide a device ID if EO can track the drone onwards

    \item Group 2 (IF1200, m600) drones are quite reasonably well detected from the start of the trajectory (with the range considered in this paper) with EO only using RPCA. Still passive RF is useful to provide a unique track ID in the 2D image plane.
\end{enumerate}

\section{Conclusions}\label{sec_conclusion}
In this work, we quantified the detection, and tracking  performance of small drones  over range with EO using advanced foreground/background separation techniques (RPCA)  as well as with combined EO and passive RF  modalities. Compared to the most existing deep learning based solutions for drone detection and identification which perform poorly  when the drones  occupy only a small number of pixels of the image plane, to the best of author's knowledge, this is the first paper to quantify the range sensitivity for small drone (Group 1, 2)  detection using EO with foreground/background separation techniques. In order to label each detected drone in the image plane, the device ID obtained using passive RF fingerprint is incorporated. For RF fingerprinting, a deep learning based model is trained to classify four drones in the AFRL ESCAPE 2021 dataset. After obtaining 3D locations with passive RF via TDOA based localization, and projecting the estimated locations to the image plane, track labels are assigned by detection-detection association. In the proposed architecture, it is sufficient to  perform passive RF based localization until RPCA based EO detections pick the corresponding tracks. This is quite appealing since passive RF may not always be available in practical counter-UAS applications. In the future, we plan to perform Monte-Carlo analysis over a range of scenarios, runs, and passes  to do track error analysis of the combined architecture with different types of drone trajectories (short, medium and long) to enhance our understanding of reliable drone detection with these two types of  modalities.  Future work also includes investigating incorporation of other modalities into the fusion framework such as acoustic and active radar  and develop a drone  intent analysis algorithm using modal-specific features as well drone kinematics estimates.

%\section{Biography}

\section{Appendix}
The augmented Lagrangian of (\ref{eq:RPCA_opt}),  $F(L,S,E,Y)$, is given by,
\begin{eqnarray*}
  F(L,S,E,Y) &=& ||L||_{\star} + \tau ||S||_1 + \lambda ||E||_F^2 \nonumber\\
   & +& {\langle Y, X-L-S-E \rangle} \nonumber\\
   &+& \frac{\beta}{2} ||X-L-S-E||_F^2
\end{eqnarray*}
where the matrix $Y$ contains the Lagrange multipliers and $\beta > 0$.
To find $L$, $\hat{L}$,  keeping $S$ and $E$ fixed:
\begin{eqnarray}
  \hat L = \underset{L}{\arg\min} \left\{ ||L||_{\star} + \frac{\beta}{2}||X-L-S-E+\frac{1}{\beta} Y||_F^2  \right\}\nonumber\\
  =\underset{L}{\arg\min}  \left\{  \frac{1}{\beta}||L||_{\star} + \frac{1}{2}||L-(X-S-E+\frac{1}{\beta} Y)||_F^2  \right\}\label{eq_Lagran}
\end{eqnarray}
The solution for (\ref{eq_Lagran}) is given by \cite{Oreifej_13}
\begin{eqnarray*}
  \hat L = D_{\frac{1}{\beta}}\left(X-S-E-\frac{1}{\beta} Y\right)
\end{eqnarray*}where $D_{\tau}(Y) = UD_{\tau}(\Sigma)V^*$, is the
singular value soft-thresholding operator where $D_{\tau}(\Sigma)=diag({(\sigma_i-\tau)_{+}})$, $(x)_{+}=max(0,x)$,   $Y=U\Sigma V^T $ is the SVD of $Y$ of rank $r$ with $\Sigma=diag({\sigma_i}_{1\leq i\leq r})$.
To find $S$, $\hat{S}$,  keeping $L$ and $E$ fixed,
\begin{eqnarray*}
  \hat S &=& \underset{S}{\arg\min} \left\{ \tau ||S||_{1} + \frac{\beta}{2}||X-L-S-E||_F^2  \right.\nonumber\\
  &+&\left. tr(Y^T(X-L-S-E))\right\}\nonumber\\
  &=&\underset{S}{\arg\min}  \left\{  \frac{\tau}{\beta}||S||_{1} + \frac{1}{2}||S-(X-L-E+\frac{1}{\beta} Y)||_F^2  \right\}
\end{eqnarray*}
for which the solution is given by soft-thresholding operator \cite{Oreifej_13}:
\begin{eqnarray*}
  \hat S =\Pi_{\frac{\tau}{\beta}}\left(X-L-E+\frac{1}{\beta} Y\right)
\end{eqnarray*}
where $\Pi_{\alpha}(x)=sign(x).max\{|x|-\alpha,0\}$. To find $E$, $\hat{E}$,  the following optimization problem is solved:
\begin{eqnarray*}
  \hat E = \underset{E}{\arg\min} \left\{ \lambda ||E||_{F}^2 + \frac{\beta}{2}||X-L-S+\frac{1}{\beta} Y||_F^2  \right\}\nonumber\\
  =\underset{E}{\arg\min} \left\{ \frac{\lambda}{\beta} ||E||_{F}^2 + \frac{1}{2}||E-(X-L-S+\frac{1}{\beta} Y)||_F^2  \right\}
\end{eqnarray*}
which reduces to
\begin{eqnarray*}
  \hat E =\frac{1}{1+2\lambda/\beta}\left(X-L-E+\frac{1}{\beta} Y\right).
\end{eqnarray*}

%%%%%%%%%%%%%%%%%%%%%%%%%%%%%%%%%%%%%%%%%%%%%%%%%%%%%%%%%%%%%%%%%%%%%%%%%%%%%%%%%%%%%%%%%%%%%%%%%%%%%%
\acknowledgments
 This work was done under an independent  research and development (IRAD) project funded by AMDS in JHU/APL. The data used in this work is part of  ESCAPE 2021 data collect of  Air Force Research Laboratory, AFRL/RI Rome, NY. Data was cleared for  public release by AFRL, Case No:  AFRL-2023-3379, on 17 July 2023. The authors would like to thank Dr Peter Zulch at AFRL/RI, Rome, NY for providing ESCAPE 2021 data and for insightful  conversations on multi-modal data fusion.

%%%%%%%%%%%%%%%%%%%%%%%%%%%%%%%%%%%%%%%%%%%%%%%%%%%%%%%%%%%%%%%%%%%%%%%%%%%%%%%%%%%%%%%%%%%%%%%%%%%%%%
%\bibliographystyle{IEEEtran}
\bibliographystyle{IEEETran}
\bibliography{IEEEabrv,reference}
\thebiography
%% This biostyle allows you to insert your photo size 1in X 1.25in
\begin{biographywithpic}
{Thakshila Wimalajeewa Wewelwala  (S'07, M'10, SM'18) }{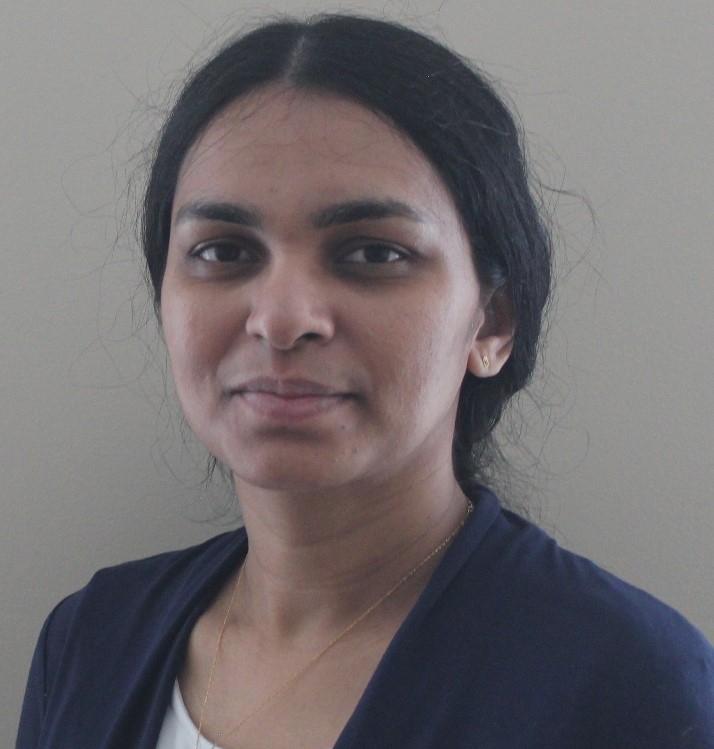}
 received her B.Sc. degree in electronic and telecommunication engineering from the University of Moratuwa, Sri Lanka, in 2004, and the M.S. and Ph.D. degrees in electrical and computer engineering from the University of New Mexico, Albuquerque, NM, USA, in 2007 and 2009, respectively. She was  with the Department of Electrical Engineering and Computer Science, Syracuse University, Syracuse, NY, as a post doctoral research  scholar from 2010 to 2012, and research faculty from 2012 to 2018. After working for BAE Systems as a Principal Research Scientist from 2018 to 2021, she joined the Johns Hopkins University Applied Physics Laboratory in 2021 as a Senior Professional Staff member. She is a Senior Member of IEEE and  the Senior Editor of the Networked Sensor Systems area of IEEE Transactions on Aerospace and Electronic Systems.  She is a recipient of  the Jean-Pierre Le Cadre Paper Award at the 21st International Conference on Information Fusion in 2018. Her research interests  span areas of statistical signal processing, multi-sensor/multi-modal data fusion and deep learning.
\end{biographywithpic}

\begin{biographywithpic}
{ Thomas W. Tedesso }{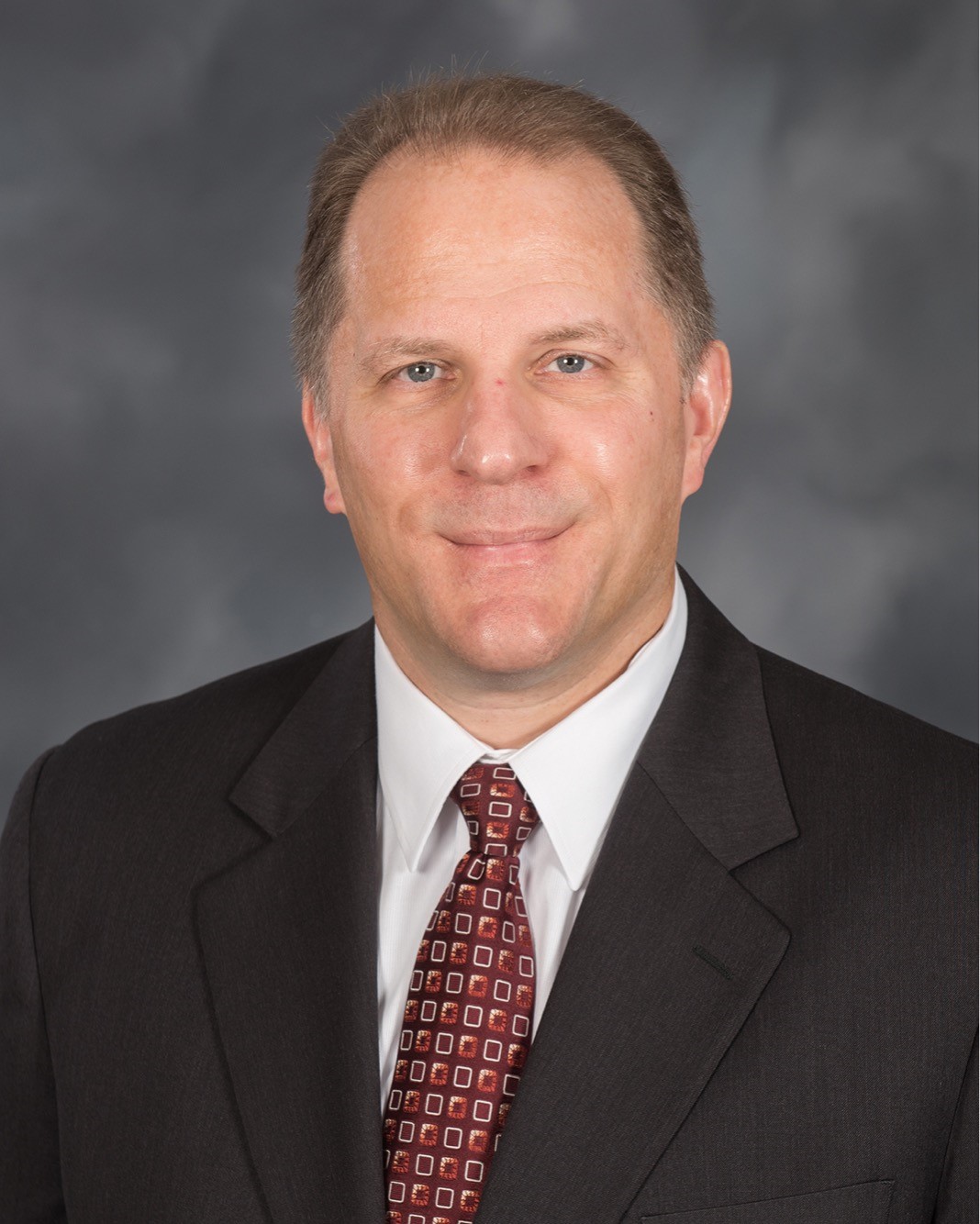}
 received the B.S. degree from the Illinois Institute of Technology, Chicago, IL, USA, in 1990, the M.S. and Ph.D. degrees from the Naval Postgraduate School, Monterey, CA, USA, in 1998 and 2013, respectively, all in electrical engineering. He served in various assignments both ashore and afloat as a surface warfare officer trained in naval nuclear propulsion and as an Assistant Professor with the United States Naval Academy, Annapolis, MD, USA. He retired from the United States Navy in November 2018.  He joined the Johns Hopkins University Applied Physics Laboratory as a Senior Professional Staff member in September 2018 and is currently serving in the role of the Assistant Group Supervisor of the Engagement Optimization Group in the Air and Missile Defense Sector.
\end{biographywithpic}

\begin{biographywithpic}
{Tony Davis}{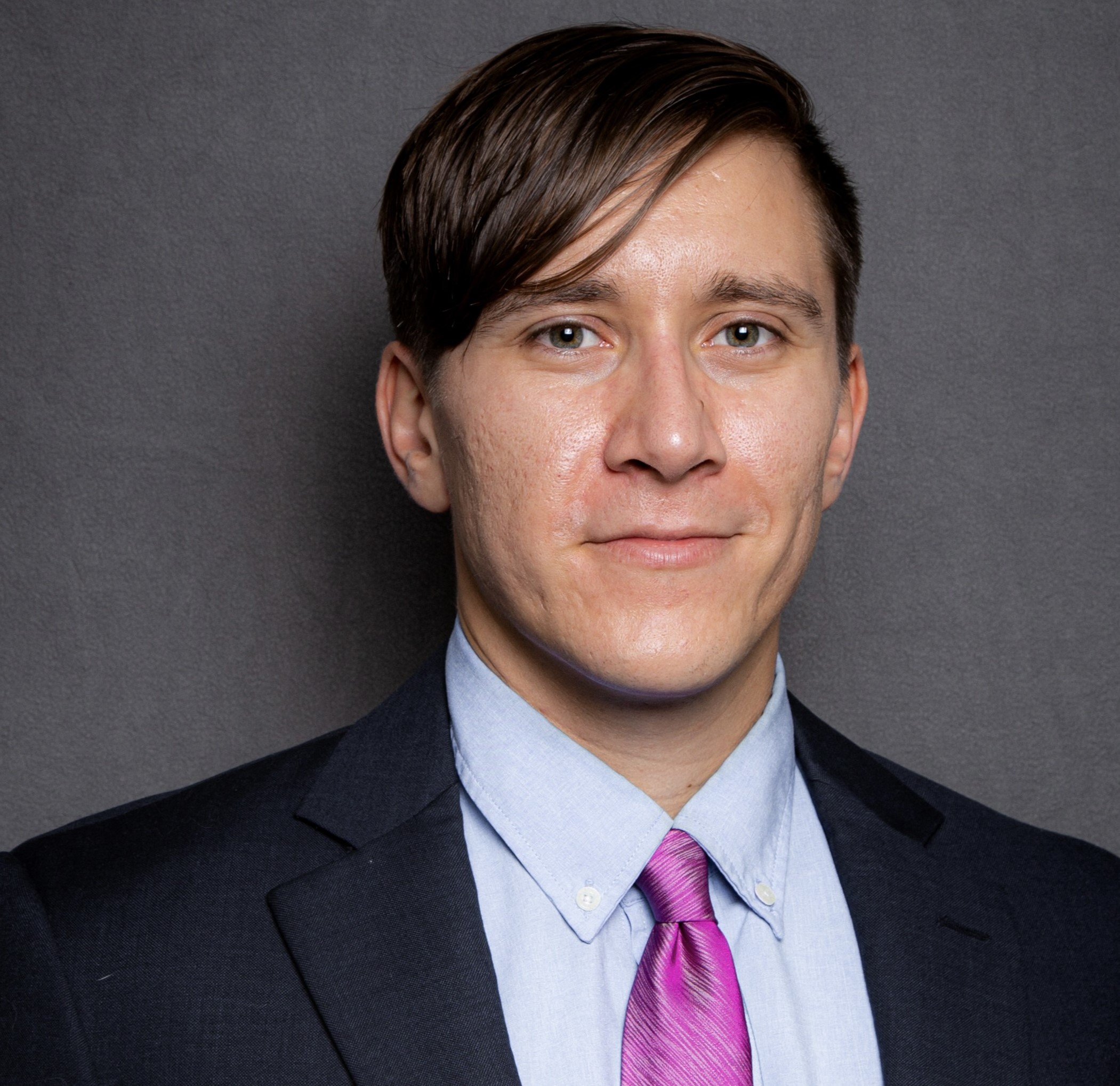}
 is a technical staff employee of the Air and Missile Defense sector of Johns Hopkins University Applied Physics Lab. He has a Master’s degree in artificial intelligence from Florida Atlantic University and has published research on optimizing drone flight planning with AI. His professional background includes machine learning software development for drug discovery at Deep Forest Sciences, mechanical engineering at Motorola Solutions, and the US Army.
\end{biographywithpic}

%%\bibliography{IEEEabr,MyBibFile}
%\begin{thebibliography}{1}
%
%\bibitem{ITAR}
%U.S. Munitions List, Sections 38 and 47(7) of the Arms Export Control Act (22 U.S.C 2778 and 2794(7).
%
%\bibitem{AeroConf}
%Aerospace Conference Web site: \underline{www.aeroconf.org}.
%
%\end{thebibliography}

%%%%%%%%%%%%%%%%%%%%%%%%%%%%%%%%%%%%%%%%%%%%%%%%%%%%%%%%%%%%%%%%%%%%%%%%%%%%%%%%%%%%%%%%%%%%%%%%%%%%%%
%\thebiography
%%% This biostyle allows you to insert your photo size 1in X 1.25in
%\begin{biographywithpic}
%{Erica Deionno}{Deionno.eps}
%received her B.S. and Ph.D degrees in chemistry from UCLA. She is currently an Assistant Principal Director in the Defense Systems Group at the Aerospace Corporation. During her 15 years at Aerospace, she has held numerous roles, including several lead positions in Aerospace's Innovation office. She also spent over 10 years conducting research in the Laboratories at Aerospace, where her work included radiation testing and modeling of emerging resistive RAM technologies and modeling space solar cell degradation.
%\end{biographywithpic}
%
%\begin{biographywithpic}
%{Jane Smith}{blankpic.eps}
%received her B.S. degree in Electrical Engineering in 1985 and a Ph.D. in Aerospacel Engineering from the Massachusettes Institute of Technology in 1990. She is currently a professor of Aerospace Engineering at the University of Nowhere. Her research includes secure communications, space cyber security, and autonomous operations.
%
%\end{biographywithpic}

\end{document}